\documentclass[twocolumn,showpacs,showkeys,prb]{revtex4}

\usepackage{graphicx}
\usepackage{dcolumn}
\usepackage{bm}
\usepackage{amssymb}

\begin{document}

\preprint{???}

\title{Quenched chirality in RbNiCl$_3$}

\author{Maikel~C.~Rheinst\"adter}
\altaffiliation[Present address: ]{Institut Laue-Langevin, 6 rue
Jules Horowitz, BP 156, 38042 Grenoble Cedex 9, France}
\email{rheinstaedter@ill.fr}  \affiliation{Institut Laue-Langevin,
6 rue Jules Horowitz, BP 156,38042 Grenoble Cedex 9,
France}\affiliation{Technische Physik, Universit\"at des
Saarlandes, PSF 1551150, 66041 Saarbr\"ucken, Germany}

\author{Garry~J.~McIntyre}
\affiliation{Institut Laue-Langevin, 6 rue Jules Horowitz, BP
156,38042 Grenoble Cedex 9, France}

\author{Mechthild~Enderle}
\affiliation{Institut Laue-Langevin, 6 rue
Jules Horowitz, BP 156,38042 Grenoble Cedex 9,
France}\affiliation{Technische Physik, Universit\"at des
Saarlandes, PSF 1551150, 66041 Saarbr\"ucken, Germany}

\date{\today}

\begin{abstract}
The critical behaviour of stacked-triangular antiferromagnets has
been intensely studied since Kawamura predicted new universality
classes for triangular and helical antiferromagnets. The new
universality classes are linked to an additional discrete degree
of freedom, chirality, which is not present on rectangular
lattices, nor in ferromagnets. However, the theoretical as well as
experimental situation is discussed controversially, and generic
scaling without universality has been proposed as an alternative
scenario. Here we present a careful investigation of the
zero-field critical behaviour of RbNiCl$_3$, a stacked-triangular
Heisenberg antiferromagnet with very small Ising anisotropy. From
linear birefringence experiments we determine the specific heat
exponent $\alpha$ as well as the critical amplitude ratio
$A^+/A^-$. Our high-resolution measurements point to a single
second order phase transition with standard Heisenberg critical
behaviour, contrary to all theoretical predictions. From a
supplementary neutron diffraction study we can exclude a
structural phase transition at T$_N$. We discuss our results in
the context of other available experimental results on RbNiCl$_3$
and related compounds. We arrive at a simple intuitive explanation
which may be relevant for other discrepancies observed in the
critical behaviour of stacked-triangular antiferromagnets. In
RbNiCl$_3$ the ordering of the chirality is suppressed by strong
spin fluctuations, yielding to a different phase diagram, as
compared to e.g.\@ CsNiCl$_3$, where the Ising anisotropy prevents
these fluctuations.
\end{abstract}

%
%
%

\maketitle

On a hexagonal lattice, an antiferromagnet can never entirely
satisfy its interactions, they will be at least partially
frustrated. A stacked set of triangular planes will nevertheless
develop long-range order for any finite inter-plane interaction.
In the perfectly isotropic case (Heisenberg antiferromagnet),
neighboring spins on a triangle compromise the antiferromagnetic
interaction by including 120~$^{\circ}$, along the hexagonal axis
the spins will be collinear. The magnetic structure is then
defined by two continuous degrees of freedom (the polar and
azimuthal angle of one chosen spin) and one additional discrete
degree of freedom, the chirality, the sense of rotation of the
spin direction on a chosen triangle. This chirality vanishes in
collinear structures, on rectangular lattices and in ferromagnets.
It is still present for easy-plane antiferromagnets, and in the
spin-flop phases of antiferromagnets with a small
Ising-anisotropy. A large family of hexagonal compounds with a
chiral degree of freedom can be described by the Hamiltonian
\begin{equation}
H=J\sum^{\mbox{intra}\atop\mbox{chain}}_{i,j}{\bf S}_i\cdot{\bf
S}_j
    +J'\sum^{\mbox{inter}\atop\mbox{chain}}_{i,k}{\bf S}_i\cdot{\bf S}_k
    -D\sum_i\left(S_i^z\right)^2.\label{Hamiltonian}
\end{equation}
Here, $J>0$ denotes the antiferromagnetic exchange interaction
between nearest neighbours along the symmetry axis, $J'>0$ the
antiferromagnetic interaction between nearest neighbours on a
triangle. The single ion anisotropy constant $D$ favors an
easy-axis ($D>0$) or plane ($D<0$). Kawamura \cite{Kaw87}
predicted that the chiral degree of freedom provokes not only a
different topology of the field-temperature phase diagrams, but
also new types of universal critical behaviour, the $n=2$ chiral
and the $n=3$ chiral universality classes. This prediction is
discussed controversially, and arguments have been given for quite
different scenarios, as e.g.\@ generic non-universal behavior
\cite{Tissier:2000}. Table~\ref{critvalues} lists Kawamura's
predictions for the critical exponents $\alpha$, $\beta$, $\gamma$
and $\delta$ and the ratio $A^+/A^-$ for antiferromagnets on
rectangular and triangular lattices as a survey.

\begin{figure} \centering
\resizebox{1.00\columnwidth}{!}{\rotatebox{0}{\includegraphics{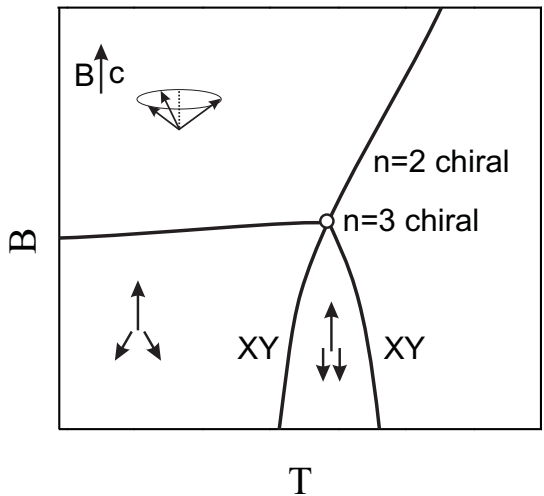}}}
\caption[]{Predicted phase diagram for ABX$_3$ with easy-axis
anisotropy. In zero magnetic field, two successive phase
transitions are expected, connected with ordering of the spin
components parallel and perpendicular to the hexagonal $c$-axis at
T$_{N1}$ and T$_{N2}$ ($<$T$_{N1}$), respectively. Both
transitions should show XY critical behavior.}
\label{pdsystematisch}
\end{figure}

ABX$_3$ compounds with easy-axis anisotropy, as CsNiCl$_3$,
RbNiCl$_3$, CsMnI$_3$ and CsNiBr$_3$ are well described by the
Hamiltonian in  Eq.~(\ref{Hamiltonian}) and have developed into
model systems for low dimensional fluctuations and ordering, see
e.g.\@ [\onlinecite{Collins:1997}] for a recent review. These
compounds show quasi one-dimensional (1D) magnetic behavior,
because the intrachain interaction $J$ is much larger than the
interchain interaction $J'$, typically $J'/J\approx 10^{-2}$.
One-dimensional short-range antiferromagnetic order within the 1D
spin chains develops below about 40~K. At lower temperatures there
is a phase transition into a three-dimensionally (3D) magnetically
ordered structure.
\begin{table*}
\begin{center}
\begin{tabular}{c|c|c|c|c|c|c}
 & & $\alpha$ & $\beta$ & $\gamma$ & $\nu$ & $A^+/A^-$ \\
\hline

 & Ising & $0.1098(29)$ & $0.325(1)$ & $1.2402(9)$ & $0.6300(8)$ &
$0.55$ \\

$\square$ & XY & $-0.0080(32)$ & $0.346(1)$ &
$1.3160(12)$ & $0.6693(10)$ & $0.99$ \\

 & Heisenberg & $-0.1160(36)$ & $0.3647(12)$ & $1.3866(12)$ &
$0.7054(11)$ & $1.36$ \\ \hline

 & n=2 chiral & $0.34(6)$ &
$0.253(10)$ & $1.13(5)$ & $0.54(2)$ & $0.36(2)$ \\

\raisebox{2.0ex}{$\bigtriangleup$} & n=3 chiral &
$0.24(8)$ & $0.30(2)$ & $1.17(7)$ & $0.59(2)$ & $0.54(2)$ \\
\end{tabular}
\caption[]{\label{critvalues}Critical exponents for
antiferromagnets on square and triangular lattices after Kawamura,
see [\onlinecite{Kaw92,kaw93,Collins:1997}] and references
therein.}
\end{center}
\end{table*}
Without an external field, Heisenberg antiferromagnets with an
Ising anisotropy on a triangular lattice undergo two successive
phase transitions, where ordering of the spin components parallel
and perpendicular to the hexagonal $c$-axis occurs at T$_{N1}$ and
T$_{N2}$ ($<$T$_{N1}$), respectively. Below T$_{N2}$, the spins
form a 120~$^{\circ}$ structure in the $ac$ plane. The predicted
B-T phase diagram is schematically shown in
Fig.~\ref{pdsystematisch}.
The two zero-field phase transitions should show 3D XY-critical
behavior \cite{Kaw87}. On a rectangular lattice, there is just one
transition with Ising-type critical behavior
\cite{Kaw92,kaw93,Collins:1997}.

In order to clarify the number of phase transitions in RbNiCl$_3$,
and their criticality, we performed linear magnetic birefringence
(LMB) experiments with a high temperature resolution to measure
the critical exponent $\alpha$ and the amplitude ratio $A^+/A^-$.
The paper is organized as follows: The properties of RbNiCl$_3$
are discussed in Sect.~\ref{RbNiCl3}. Experimental details of the
birefringence set-up are presented in Sect.~\ref{Experimental},
the LMB results and the outcome of a supplementary neutron
diffraction study are shown and discussed in Sect.~\ref{Results}.
The anomalous behavior of RbNiCl$_3$ as compared to other members
of the above mentioned ABX$_3$ family, is discussed in the
discussion in Sect.~\ref{Discussion}.

\section{RbNiCl$_3$\label{RbNiCl3}}
\begin{figure}
\centering
\resizebox{0.75\columnwidth}{!}{\rotatebox{0}{\includegraphics{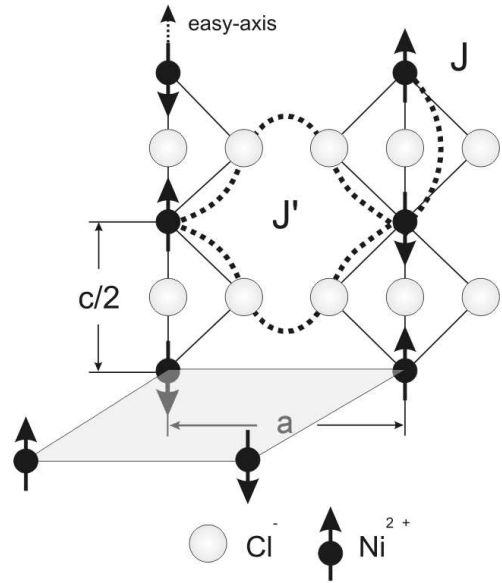}}}
\caption{In RbNiCl$_3$ magnetic exchange $J$ along the easy-axis
is two orders of magnitude larger than exchange in the basal plane
$J'$, which involves two Cl$^-$-ions (as compared to one along
$c$).} \label{austauschpfade}
\end{figure}

RbNiCl$_3$ is a quasi 1D S=1 Heisenberg antiferromagnet with a
weak Ising anisotropy on a triangular lattice (hexagonal space
group P6$_3$/mmc). As in other members of the ABX$_3$ family,
CsNiCl$_3$, CsMnI$_3$, CsNiBr$_3$ and RbNiBr$_3$, the magnetic
Ni$^{2+}$-ions form strongly coupled chains along the
crystallographic $c$-axis. The chains are characterized by an
intrachain exchange parameter $J$, which is much larger than the
interchain exchange parameter $J'$ because magnetic exchange in
the basal plane is mediated via two X-ions compared to only one
along $c$, as pictured in Fig.~\ref{austauschpfade}. $J'/J=0.38
\mbox{K}/23.8 \mbox{K}=1.6\cdot 10^{-2}$ in RbNiCl$_3$
\cite{Nak92}. The magnetic behavior therefore is quasi 1D.
At T$_N\simeq$ 11~K, there is a phase transition into a 3D
magnetically ordered structure.

\begin{table*} {\small
\begin{center}
\begin{tabular}{c|c|c|c|c}
Technique& Ref.\@ & T$_N$ (K) & $\alpha$ & $\beta$\\\hline

neutron diffraction & [\onlinecite{Yel72}] & 11.15 & &
$\beta=0.30\pm 0.01$
\\

 & [\onlinecite{Ooh91b}] &
11.11,11.25 & & $\beta_{\parallel,\perp}=0.27\pm 0.01,0.28\pm
0.01$  \\\hline
LMB & [\onlinecite{Iio80}] & 11 & &  \\

 & [\onlinecite{Ooh91}] & 11.3 & $0.06\pm0.04$ &  \\

 & [\onlinecite{Ooh94}] & 11 & &  \\ \hline

NMR  & [\onlinecite{Mun95}] & 11.18,11.36 (?)& &  \\
\hline

susceptibility, torque & [\onlinecite{Tanaka:1992,Tan93}] & 11.38
& &  \\ \hline

magnetization, susceptibility & [\onlinecite{Johnson:1979}] & 11 & &  \\
\hline

thermal expansion & [\onlinecite{Rayne:1981}] & 11.2 & &  \\
\hline

specific heat capacity & [\onlinecite{Rayne:1981,Collocott:1987}] & 11.0 & &  \\

\end{tabular}
\caption[]{Reported results for the zero-field phase transition in
RbNiCl$_3$. Values for the relation $A^{+}/A^{-}$ are not given in
the references.}\label{Literatur}
\end{center}
}
\end{table*}
Magnetic ordering in RbNiCl$_3$ can be discussed in the context of
other members of the ABX$_3$ family. In e.g.\@ CsNiCl$_3$, two
successive phase transitions are found in neutron scattering,
magnetic birefringence and specific heat capacity experiments and
display 3D XY-critical behavior with the corresponding critical
exponents \cite{Beckmann:1983,Collins:1997}, as predicted by
Kawamura. For RbNiCl$_3$, with most experimental techniques just
one transition is observed. The criticality of this transition is
not clear: Different methods obtained disagreeing values of the
critical exponents and accordingly different universality classes
have been proposed for the transition. None of the experimentally
determined values agrees with the prediction for the 3D XY class.
Table \ref{Literatur} summarizes experimental techniques and the
values determined for T$_N$, $\alpha$ and $\beta$, as found in
literature. Apart from a neutron scattering study by Oohara {\em
et al.\@} \cite{Ooh91b}, all measuring techniques report only one
phase transition. The temperature resolution in all experiments
was better than 0.02~K, considerably smaller than 0.15~K, claimed
as the distance between T$_{N1}$ and T$_{N2}$ in the neutron
scattering study. The anomalies in all techniques (except
[\onlinecite{Ooh91b}]) appear very sharp while the overlap of two
close lying divergences would lead to a rounded and broad anomaly.
Furthermore, the measured critical exponents do not coincide with
the predicted 3D XY-critical behavior. If the two transitions
would fall together at the same temperature, the transition from
the paramagnetic directly into the chiral ordered state should
show n=3 chiral exponents, against the experimental evidence.

RbNiCl$_3$ has a very small Ising anisotropy $D$, as compared to
other members of the ABX$_3$ family. We argue that the pronounced
Heisenberg character plays the key role for the understanding of
phase transitions and criticality in RbNiCl$_3$. In the next
section, we present and discuss the results of our high resolution
LMB experiments.

\section{Experimental\label{Experimental}}
The linear birefringence $n_{ac}=n_c-n_a$ has been measured using
a S\'{e}narmont set-up \cite{Sen1840,Belanger:1984} with a He-Ne
laser at $\lambda$=632.8 nm. Before and behind the sample,
apertures with a diameter of 0.3 mm were installed. The
sensitivity of the S\'{e}narmont set-up was increased by
modulating the incoming polarisation with 50~kHz and lock-in
detection of the intensity. Single crystals of RbNiCl$_3$ were
grown by the Bridgeman method. The slightly hygroscopic samples
were prepared by cleaving in a glovebox under He-atmosphere. The
natural cleavage planes contain the $c$-axis, and correspond
probably to $\{10\bar{1}0\}$. The typical sample size was 4x4x1.5
mm$^3$. The cleft samples were used without further polishing and
were mounted stress-free in an optical $^4$He continuous flow
cryostat with a temperature stability of 0.001~K. The sample
temperature was measured with a Cernox semiconductor thermometer
in lock-in technique with a relative accuracy of 0.001~K.

Under certain conditions, the derivative $dn_{ac}/dT$ is
proportional to the magnetic part of the specific heat capacity ,
see e.g.\@ [\onlinecite{Ferre:1984}] and references therein. This
relation is in particular valid close to the phase transitions of
the antiferromagnetic triangular ABX$_3$ compounds with and
without easy-axis anisotropy, as CsNiCl$_3$ and RbNiCl$_3$. In the
temperature range of the phase transition in RbNiCl$_3$ at
$T_N\approx 11$~K, the specific heat capacity is already dominated
by contributions of the crystal lattice. The critical properties
of the magnetic specific heat are therefore difficult to measure
in a standard specific heat capacity setup. Here the
birefringence is an elegant way to determine the critical exponent
$\alpha$ as well as the amplitude ratio $A^{+}/A^{-}$ of the
critical part of the specific heat capacity above and below the
phase transition.

\section{Results\label{Results}}
\begin{figure}
\centering
\resizebox{1.00\columnwidth}{!}{\rotatebox{0}{\includegraphics{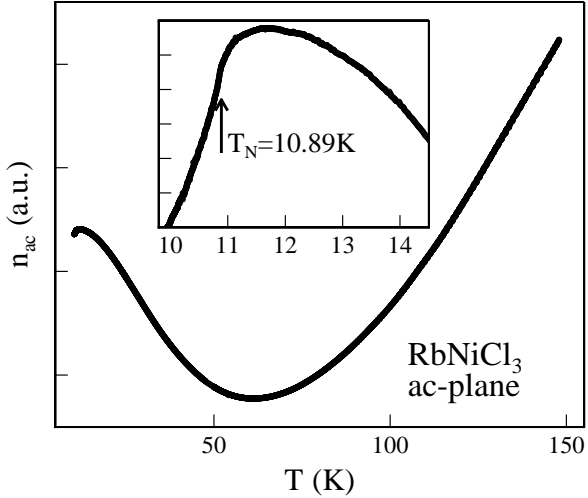}}}
\caption[]{Temperature dependence of the birefringence
$n_{ac}=n_c-n_a$ over a broad temperature range. The inset shows
the range of the phase transition in magnification. The transition
is marked by an arrow.} \label{deltan}
\end{figure}
Figure \ref{deltan} shows the temperature dependence of $n_{ac}$
over a broad temperature range. At high temperatures, $n_{ac}$
linearly decreases with lowering temperature. Below about T=70~K
there is distinct deviation from linear behavior due to the onset
of short ranged 1D correlations along the Ni-chains \cite{Iio80}.
The inset in Fig.~\ref{deltan} shows the temperature range of the
3D phase transition in magnification. The onset of 3D correlations
close to T$_N$=10.89~K, which finally leads to a 3D magnetically
ordered structure, is indicated by the drop of the birefringence
below 11~K.

The derivative of the birefringence with respect to temperature is
described by a power law:
\begin{equation}
\frac{dn_{ac}}{dT}=A^{\pm}\left|\frac{T-T_N}{T_N}\right|^{-\alpha^{\pm}}.
\label{criteq}\end{equation}
\begin{figure} \centering
\resizebox{1.00\columnwidth}{!}{\rotatebox{0}{\includegraphics{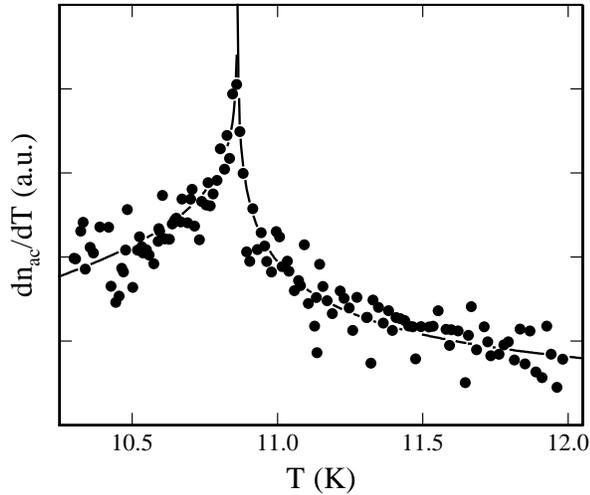}}}
\caption[]{Temperature-derivative of the critical part of the
birefringence, $dn_{ac}/dT$, which is proportional to the magnetic
specific heat. The solid line is the resulting fit with
Eq.~(\ref{criteq}).} \label{difplot}
\end{figure}
Figure~\ref{difplot} shows $dn_{ac}/dT$, the solid line is a fit
after Eq.~(\ref{criteq}). The noncritical contribution due to 1D
correlations and lattice natural birefringence was taken into
account by a polynomial of the form $a+bT+cT^2+dT^3+eT^4$ which
was subtracted from the data. We observe only one transition as
the fitted transition temperatures for the range below and above
T$_N$ perfectly coincide. The good temperature resolution allows
to measure as close to the phase transition as $10^{-4}$ in
reduced temperature, considerably closer than all previous
experiments. Even if two different transition temperatures were
allowed for the high and the low temperature side, they converge
to a single one in the fit. We do not observe any signs of
crossover effects.
\begin{figure}
\centering
\resizebox{0.80\columnwidth}{!}{\rotatebox{0}{\includegraphics{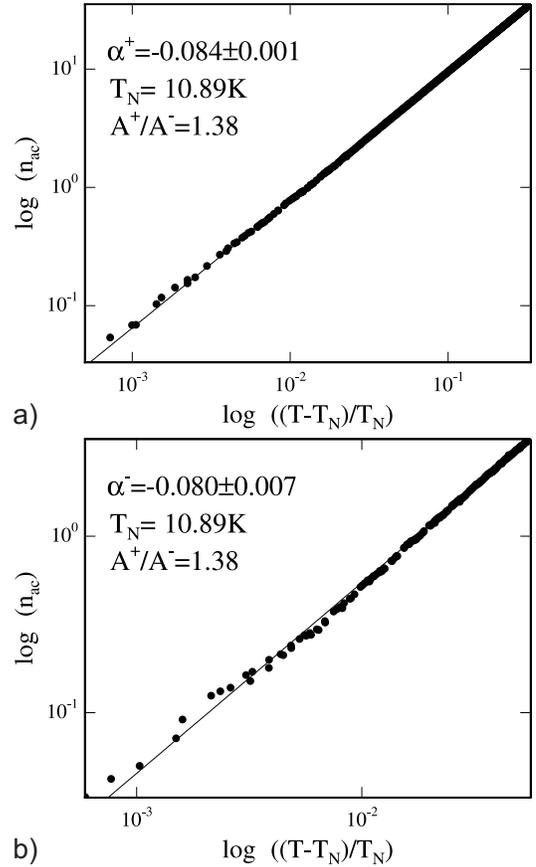}}}
\caption[]{Log-log plots of the critical part of the birefringence
$n_{ac}$ vs. reduced temperature $|t|=|(T-T_N)/T_N|$ for (a)
T$>$T$_N$ and (b) T$<$T$_N$. Solid lines are fits with
Eq.~(\ref{criteq}), the fitted values for $\alpha^{\pm}$, T$_N$
and the ratio A$^+$/A$^-$ are given in the figure.} \label{logpic}
\end{figure}
To check the quality of the fits, Fig.~\ref{logpic} shows log-log
plots of the critical part of $n_{ac}$ vs. reduced temperature
$|t|=|(T-T_N)/T_N|$ for T$\lessgtr$T$_N$ in the range close to the
phase transition. T$_N$ is determined to T$_N=10.888\pm 0.001$~K,
the values for $\alpha$ from the high and the low temperature side
to $\alpha^{+}=-0.084\pm 0.001$ and $\alpha^{-}=-0.080\pm 0.007$.
The ratio $A^+/A^-$ is obtained to $1.38\pm 0.07$. Comparing these
values with those from Tab.~\ref{critvalues}, the determined
critical exponent and $A^+/A^-$ agree remarkably well with those
of a Heisenberg antiferromagnet on the rectangular lattice. Chiral
or XY-behavior, as predicted in the chiral theory, can be
excluded. Two close lying successive phase transitions that would
lead to a rounded anomaly in the measurements can obviously be
excluded by our measurements in Figs.~\ref{difplot} and
\ref{logpic}.

The unusual behavior of RbNiCl$_3$ might be explained by a lift of
degeneracy of the magnetic exchange interactions in the hexagonal
basal plane. This scenario has been discussed for other ABX$_3$
compounds in e.g.~[\onlinecite{Collins:1997,Zhitomirsky:1995}].
Considering the crystal structure of RbNiCl$_3$, as pictured in
Fig.~\ref{austauschpfade}, a lift of degeneracy is inseparable
from changes in the crystal lattice.
We therefore carried out supplementary single-crystal neutron
diffraction experiments at the new Vivaldi Laue-diffractometer at
the high flux reactor of the ILL in Grenoble, France, to detect a
possible change in the lattice symmetry below T$_N$. Vivaldi's
large image plate thereby allows to survey large areas of
reciprocal space to detect possible superlattice reflections in
the ordered phase.
\begin{figure}
\centering
\resizebox{1.00\columnwidth}{!}{\rotatebox{0}{\includegraphics{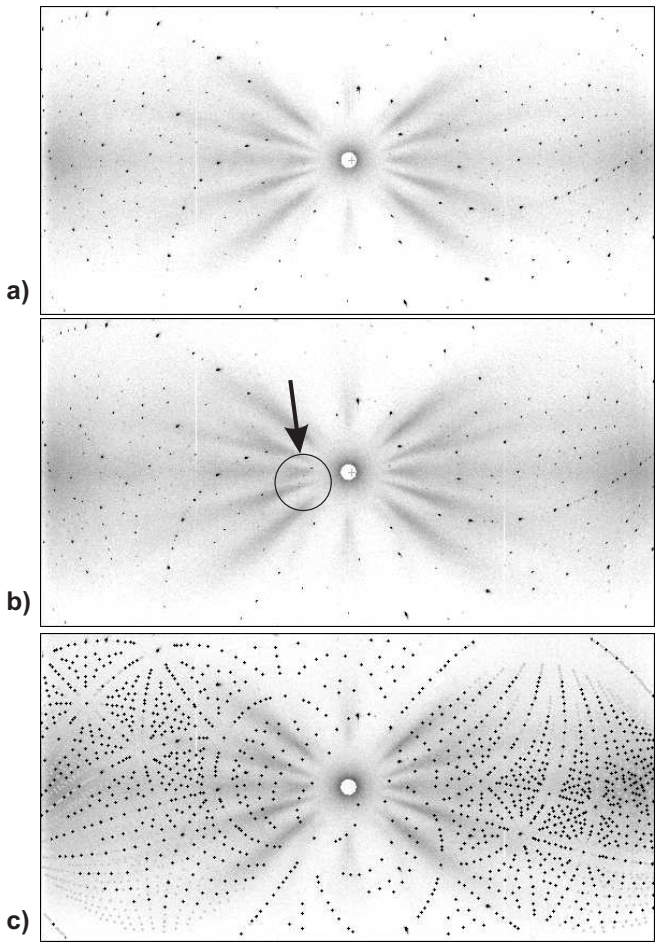}}}
\caption[]{Laue-photographs taken at different temperatures. (a)
T=20~K, above phase transition; (b) T=2~K, in the magnetically
ordered phase (some of the magnetic superlattice reflections are
marked by the arrow); (c) T=2~K with predicted Laue-pattern.}
\label{vivaldibild.eps}
\end{figure}
Typical sample crystals of about 1x1x2 mm$^3$ were mounted in a
helium orange cryostat. We took exposures at T=20~K, in the
paramagnetic, and in the ordered phase, at 2~K. The corresponding
Laue patterns are shown in Fig.~\ref{vivaldibild.eps}. The
reflections of the T=20~K exposure in Fig.~\ref{vivaldibild.eps}
(a) could be indexed by a primitive hexagonal cell with lattice
parameters $a$=6.93\AA\ and $c$=5.89\AA. The reflections in the
magnetically ordered phase can be described in terms of a tripled
hexagonal cell $(a\sqrt 3,a\sqrt 3,c)$. We could not detect any
splitting of the reflections below the phase transition within the
experimental angular resolution of 10 ' nor the appearance of
additional superlattice reflections which are not indexed by the
tripled hexagonal cell. Even a small orthorhombic or monoclinic
distortion would lead to the appearance of Bragg peaks at former
forbidden positions and should have been detected. Our
measurements therefore confirm the previous results by Yelon and
Cox \cite{Yel72}. Moreover, the zero-field birefringence $n_{ab}$
in the hexagonal basal plane \cite{Ooh91} vanishes, which
independently excludes any orthorhombic or monoclinic distortion
in the ordered phase.

\section{Discussion\label{Discussion}}
The question arises, why RbNiCl$_3$ - which orders into the same
magnetic structure as CsNiCl$_3$ - does not show two successive
phase transitions and the predicted chiral critical behavior. Two
successive phase transitions can be excluded from our high
resolution birefringence measurements as well as from most of the
previously reported experimental results.
Close lying divergences due to two close lying phase transition
should lead to a rounded anomaly in the measurements. But even in
the high resolved data of Fig.~\ref{difplot}, the anomaly remains
sharp and pronounced confirming the single phase transition
observed in a previous LMB study \cite{Ooh91} and other techniques
(see the listing in Tab~\ref{Literatur}). The critical exponent
$\alpha$ and the ratio $A^+/A^-$ correspond to conventional
Heisenberg critical behavior and therefore point to a disordered
chirality below T$_N$. A vanishing chirality due to a collinear
structure can be excluded from the structural data. If there was
only one transition, connected with ordering of the spin
components parallel and perpendicular to the 1D axis but no static
ordering of the chirality, the corresponding transition should
indeed show conventional Heisenberg critical behavior like for
antiferromagnets on rectangular lattices. We argue in the
following that spin fluctuations suppress long ranged chiral order
in RbNiCl$_3$ below T$_N$.

\begin{figure} \centering
\resizebox{0.65\columnwidth}{!}{\rotatebox{0}{\includegraphics{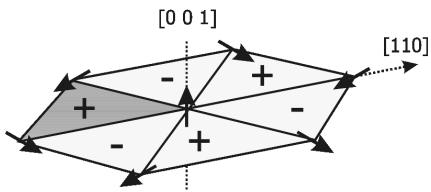}}}
\caption{In the magnetically ordered phase of RbNiCl$_3$ as
proposed in literature, 2/3 of the spins are canted away from $c$
in the [110] direction. The chirality $\vec{\kappa}$ changes sign
from one triangle to the neighboring
triangle.}\label{chiralitypattern}
\end{figure}

The chirality, which basically takes into account the sense of
rotation of the spin direction on a chosen triangle, is defined as
\cite{Kaw88b}:
\begin{equation}
\vec{\kappa}=\frac{2}{3\sqrt{3}}\left({\bf S_i}\times{\bf
S_j}+{\bf S_j}\times{\bf S_k}+{\bf S_k}\times{\bf S_i}\right).
 \end{equation}
Figure~\ref{chiralitypattern} shows the ordered spin structure of
RbNiCl$_3$ in the hexagonal basal plane, as proposed in the
literature. The spins lie in a $[001][110]$ plane with 2/3 of the
spins canted away from $c$ by an angle $\theta$. $\theta$ depends
on the ratio $D/J'$ and is determined to $\theta$=57.5~$^{\circ}$
\cite{Yel72} in RbNiCl$_3$, very close to the ideal value of
60~$^{\circ}$. In this model, the chirality $\vec{\kappa}$ is long
ranged ordered and changes sign from one to the neighboring
triangle, respectively. Anti-phase domains of the chirality
contribute equally in a neutron scattering experiment.
Oohara and Iio investigated the RbNi$_{1-x}$Co$_x$Cl$_3$ system
\cite{Tan93} with LMB. By replacing Ni$^{2+}$ by Co$^{2+}$, the
magnitude of the Ising anisotropy $D$, which is very small in pure
RbNiCl$_3$ (70 \% that of CsNiCl$_3$), can gradually be increased.
With increasing $D$, two anomalies become visible in the LMB
experiments and the distance T$_{N1}$-T$_{N2}$ increases. The
latter study clearly shows that the small Ising anisotropy $D$
plays the crucial role for the understanding of criticality and
phase transitions in RbNiCl$_3$. It also proves that LMB is
capable to detect the upper transition, if it exists.

The anisotropy $D$ confines the 120~$^{\circ}$ spin structure in
the $ac$ plane. Depending on the ratio $D/J'$, the structure might
exhibit an additional degree of freedom connected with the
rotation of the 120~$^{\circ}$ structure in the $ac$ plane. This
{\em quasidegeneracy} has been predicted \cite{Miyashita:1986} and
experimental evidence was found for the case of CsNiCl$_3$
\cite{Maegawa:1988}. The energy barrier for a rotation of the
spin-star in the $ac$ plane is of the order of $D(D/6J')^2$
[\onlinecite{Ooh94}]. Miyashita \cite{Miyashita:1986} suggested,
this quasidegeneracy exists, if $(D/J')<1$ ($D/J'=0.06$ in
RbNiCl$_3$). Even though $D_{RbNiCl_3}= 0.7D_{CsNiCl_3}$,
$D(D/6J')^2$ for RbNiCl$_3$ is just 7 \% of that of CsNiCl$_3$;
the quasidegeneracy should therefore be strongly enhanced in the
paramagnetic phase of pure RbNiCl$_3$.

\begin{figure} \centering
\resizebox{1.00\columnwidth}{!}{\rotatebox{0}{\includegraphics{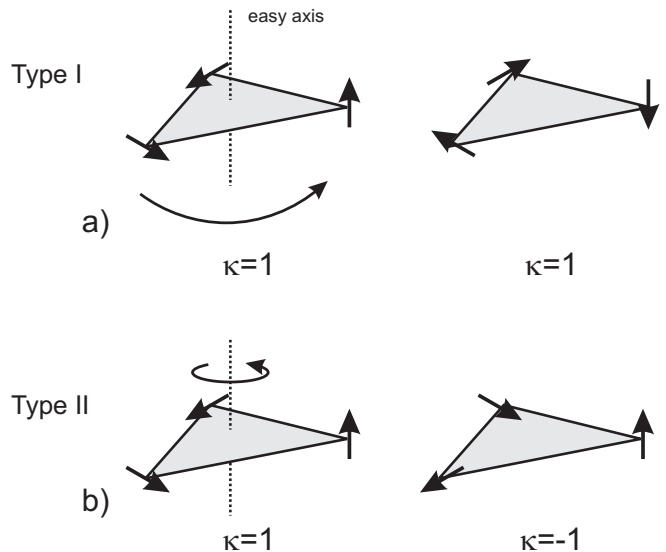}}}
\caption{Fluctuations of the triangle marked in
Fig.~\ref{chiralitypattern}: (a) Type I: Rotations about an axis
parallel to the vector of chirality $\vec{\kappa}$ preserve the
chirality. (b) Type II: Fluctuations around the easy-axis,
perpendicular to $\vec{\kappa}$, change sign of the chirality.}
\label{mercedessternrelax}
\end{figure}

NMR and measurements of the specific heat capacity (see
Tab.~\ref{Literatur}) give evidence for strong spin fluctuations
also in the ordered phase of RbNiCl$_3$.
Figure~\ref{mercedessternrelax} schematically shows the two basic
spin relaxation mechanisms. Type I fluctuations are rotations of
the spin-star around an axis perpendicular to the spin plane,
i.e.\@  parallel to the chirality vector $\vec{\kappa}$. This is
the quasidegeneracy that has been discussed above. As indicated in
the figure, these fluctuations preserve the chirality of the
triangle; $\vec{\kappa}$ can still show long ranged order.
All fluctuations with axis of rotation perpendicular to
$\vec{\kappa}$ (Type II fluctuations) change the sign of
$\vec{\kappa}$. If these fluctuations occur incoherently,
$\vec{\kappa}$ cannot order. The phase transition should be of
conventional type, as suggested by the LMB experiment.

Type II fluctuations seem not to depend directly on the Ising
anisotropy $D$ because the canting angle of the respective spins
does not change during the rotation. Their incoherent occurrence
in the ordered structure, however, may be emphasized by the
presence of Type I fluctuations. when the Ising anisotropy is
enlarged In CsNiCl$_3$ or in the RbNi$_{1-x}$Co$_x$Cl system, the
contribution of Type II fluctuations is obviously negligible, as
these compounds show chiral ordering as predicted by theory. This
seems to imply that Type II fluctuations play a major role only
when Type I fluctuations are already strongly enhanced (as in pure
RbNiCl$_3$).

The basic idea of fluctuations which on the one hand preserve
(Type I) and on the other hand suppress (Type II) long ranged
chiral order seems to account well for phase transitions and
critical behavior observed in RbNiCl$_3$. The separate ordering of
the spin components parallel to the 1D axis, is presumably
suppressed by Type I fluctuations; Type II fluctuations do not
affect the projection of the magnetic moment onto the $c$-axis.
But fluctuations of Type II might suppress ordering of the
chirality $\vec{\kappa}$ at the phase transition T$_{N}$ where the
magnetic moment shows 3D ordering (whereas Type I fluctuations
have no effect on the sign of $\vec{\kappa}$). If both types of
fluctuations are strongly enhanced, we imagine that the domain
walls between chirality domains of opposite sign move freely
through the otherwise magnetically long-range ordered structure.
If the chirality domain walls in the ordered phase behave
liquid-like,
the transition should show conventional Heisenberg critical
behavior, as it is observed in the LMB experiment. In this
language, the chirality domain walls in CsNiCl$_3$ or CsMnBr$_3$
are quasi-static. This liquid-like behavior of the domain walls,
which leads to a different phase diagram and different critical
behavior, as compared to other members of the ABX$_3$ family, must
crucially depend on the almost perfect Heisenberg character of
RbNiCl$_3$.

\section{Conclusions\label{Conclusions}}
We present a linear magnetic birefringence study in RbNiCl$_3$.
Our high resolution determination of the critical parameters
$\alpha$ and the amplitude ratio $A^+/A^-$ show conventional
Heisenberg critical behavior like antiferromagnets on rectangular
lattices (which have no ordered chirality) as opposed to
theoretical predictions. There is just one phase transition in
RbNiCl$_3$. From a neutron diffraction study we can exclude a
structural phase transition and a lift of the degeneracy of the
magnetic exchange interactions in the basal plane at T$_N$. We
discuss RbNiCl$_3$ in the framework of previous experimental and
theoretical results and other members of the ABX$_3$ family. We
finally argue that spin fluctuations lead to the unusual behavior
of RbNiCl$_3$. A separate phase transition of the spin component
parallel to the easy-axis might be suppressed by spin fluctuations
with axis of rotation parallel to the chirality vector
$\vec{\kappa}$ (Type I fluctuations). Fluctuations of Type II,
with axis of rotation perpendicular to $\vec{\kappa}$, presumably
suppress long ranged order of the chirality $\vec{\kappa}$ below
T$_N$. The resulting single phase transition shows conventional
Heisenberg critical behavior, as evidenced by the critical
exponents and phase transitions observed in the LMB experiment.

\acknowledgments We are indebted to and thank K.~Knorr for
hospitality and fruitful discussions. This work has been partially
funded by the Universit\"at des Saarlandes, Saarbr\"ucken,
Germany. We thank H.~Tanaka for providing us with the high quality
samples.

\bibliography{Rbnicl3}

\end{document}